# Automated Program Analysis for Novice Programmers[1]

**Tim Blok and Ansgar Fehnker**

Faculty of Electrical Engineering, Mathematics and Computer Science,
University of Twente, Enschede, the Netherlands

*Abstract*

*This paper describes how to adapt a static code analyzer to help novice programmers. Current analyzers have been built to give feedback to experienced programmers who build new applications or systems. The type of feedback and the type of analysis of these tools focusses on mistakes that are relevant within that context, and help with debugging the system. When teaching novice programmers this type of advice is often not particularly useful. It would be instead more useful to use these techniques to find problem in the understanding of students of important programming concepts.*

*This paper first explores in what respect static analyzers support the learning and teaching of programming can be implemented based on existing static analysis technology. It presents an extension to static analyzer PMD was made so that feedback messages appear which are easier to understand for novice programmers. To answer the question if these techniques are able to find conceptual mistakes that are characteristic for novice programmers make, we ran it over a number of student projects, and compared these results with publicly available mature software projects.*

---

## 1. Introduction

Professional software development teams are using a range of tools to detect and correct mistakes in their programs. It is common to distinguish between dynamic and static analysis tools. Dynamic code analyzers look at code while it is executing, often by running a given set of tests, or test that have been generated. Static code analyzers check for programming errors in code by automated inspection of the code. This includes simple errors, like violations of programming style, or uninitialized variables to, serious errors and often difficult to detect error, such as memory leaks, race conditions, or security vulnerabilities.

This paper describes how to adapt an off-the-shelf static code analysis tools to support teaching of novice students. This requires first an analysis to what extend the different context in which it is used changes the requirements for the tool. Static analysis tools such as Coverity and FindBugs are meant for experienced developers, who know how to program. They do not look for errors that experienced programmers rarely if ever make, and focus instead on errors that even experienced programmers find difficult to debug for (Bessey, et al., 2010) (Hovemeyer & Pugh, 2004).

Another crucial difference is that the tools for development teams are built to aid with debugging a software system. An error that is found statically in the code may produce a run-time error in de production code, and the tools help with correcting those. In a learning context errors usually point to lack of understanding of important an concept, from basic concepts, such as the difference between a variable and a value, to more advanced concepts such as encapsulation. The foremost aim is not to correct the program, but to help students and teachers to identify misunderstood concepts, such that these can be addressed.

Other consequences arise from changing the context in which the tool is used. Reported errors may relate to the concepts which the students does not know, and a student students may get stuck trying to address a warning that is not understood. In the process they often introduce further errors that make the warning go away, but only since the changes obfuscated the initial error. The tool should recognize that these are not the problem itself.

Finally, there is also the fact that software development team like to see coding guidelines enforced that make sense in that context, for example compliance with industry standards, or making their code analyzable for other tools, or ensuring consistency among different platforms. For student these warning can often be overwhelming, confusing, or a nuisance.

This paper discusses how to adapt PMD[2] with custom rule to suit the needs of novice programmers in JAVA. The benefits of using an off-the-shelf tool is that we have access to the underlying analysis technology that has been proven its use in practice. It also means that the tool can be readily used in an Integrated Development Environment, or even and automated build system. The extension of PMD aims at both students and their teachers. Students see immediately see what they did wrong, while it helps teacher to speed up the process of looking through assignments and give meaningful feedback. It would also standardize to some extend the issues different markers look for.

---

[2] Available at https://pmd.github.io/

Dynamic tools have been incorporated in numerous teaching environments, most prominently in BOSS (Joy, Griffiths, & Russell, 2005). FrenchPress is uses static analysis to provide feedback to students, however is intentionally not built on an existing tool such as PMD (Blau, Eliot, & Moss, 2015). It implements a number of tests similar to those presented in this paper, but does not use the full range that analysis static tools offer. Sen reviews the usage of code analysis in programming class (Sen, 2014), and mentions that custom rules can help in the classroom, however does not explore the rules themselves. Machine learning techniques are used in (Srikant & Aggarwal, 2014) to analyze student code, but the focus lies on automated grading instead of giving feedback that relates error to misunderstood concepts.

The next section will discuss the challenges of static analysis for novice programmers, and the type of behavior we might observe. Section 3 will discuss custom rules, and Section 4 an experimental evaluation, that compares student projects with publicly available software projects, that were presumably developed by experts. These results confirm that those rules are effective in detecting error made by novices.

## 2. Challenges for Static Code Analysis for Novice Programmers

Figure 1 depicts an example of code that is typical of student code. It points to a student who fails to understand the difference between character variables and character literals. The loop is meant to check if the reply is the character 'Y', but instead the student compares it to the (undefined variable) Y. The program would initially not compile, and the compiler would

```
char   reply;
char Y;

do{
  System.out.println("Do  you   want   to   continue?");
  reply  =  (char)  System.in.read();
} while(reply != Y);
                ^
Variable 'Y' might not have been initialized
```

*Figure 1: Typical static analysis tool feedback*

issue an error message about an undefined variable `Y`. Many novice programmers would address this error by introducing a variable `char Y`. This results in the code in Figure 1.

Some compilers, and most static analyzers, would warn that variable `Y` has not been initialized. Many students would then try to address this warning, and initialize the variable, such as `char y = 0;` or `char y = '0';`, even though this does nothing to address the actual error. This will remove the error warning, but only conceal the bigger problem, namely that the student confuses the literal 'Y' with a variable 'Y'. In contrast, when an experienced programmer gets the first warning about variable '`Y`' being undefined, they will immediately realize that they simply forgot the quotes around the `Y`.

Another example is the following one-liner:

```
if(string1 != null || string1.equals(string2)){...}
```

The current feedback of PMD will state that the variable '`string1`' will always be null at the second occurrence of '`string1`' in this line. To understand the message the student has to understand the evaluation order of short circuit operators || and &&. The right hand side of the operator || will only be evaluated, if the left hand side evaluates to false. The left hand side in this example if only false if `string1 != null` is false, thus if the string is null. Hence, the warning that `string1` will always be null when `string1.equals(string2)` is evaluated. The actual problem has nothing to do with the evaluation order of short circuit operators. The student simply mixed up || and &&, a very common mistake by novice programmers. The feedback should ask the student if he accidentally mixed up the operators, and intended to use &&.

A list of 20 errors, which are often made by beginning programmers was created and discussed by Hristova (Hristova, 2003). We use this list to address the following questions: Which of these error can static analysis tools find? What kind of information has to be relayed to a novice programmer to know what they did wrong? Can this be related to concepts that the students may misunderstand? How effective are the added and modified rules in finding issues that are characteristic of novice code?

As mentioned before, PMD has been chosen as the tool to extend. However, this is far from the only tool available for Java code analysis. The three most popular tools for Java are FindBugs, PMD and Checkstyle. Checkstyle is mostly concerned about correcting certain code styles. FindBugs is a tool that looks at the java byte code. This is very useful for detecting serious coding errors, it limits it use to define the appropriate rules at the level of the uncompiled syntax. PMD is a good hybrid between the two previous tools. It utilizes a generated abstract syntax tree (AST) from the source code, and then uses code patterns to identify bad practices. All tools have in common that code that cannot be parsed can also not be checked by tools.

> A. Confusing the assignment operator (=) with the comparison operator (==).
> B. Use of == instead of .equals to compare strings.
> C. Unbalanced parentheses, curly brackets, square brackets and quotation marks, or using these different symbols interchangeably.
> D. Confusing short-circuit evaluators (&& and ||) with conventional logical operators (& and j).
> E. Incorrect use of semi-colon after an if, while or for statement.
> F. Wrong separators in for loops (using commas instead of semi-colons).
> G. Inserting the condition of an if statement within curly brackets instead of parentheses.
> H. Using keywords as method names or variable names.
> I. Invoking methods with wrong arguments (e.g. wrong types).
> J. Forgetting parentheses after a method call.
> K. Incorrect semicolon at the end of a method header.
> L. Getting greater than or equal/less than or equal wrong, i.e. using => instead of >=
> M. Trying to invoke a non-static method as if it was static.
> N. A method that has a non-void return type is called and its return value ignored/discarded.
> O. Control ow can reach end of non-void method without returning.
> P. Including the types of parameters when invoking a method.
> Q. Incompatible types between method return and type of variable that the value is assigned to.
> R. Class claims to implement an interface, but does not implement all the required methods.
> S. Confusing character variables as literals
> T. Null check followed by ||
> U. Many if/else checks on the same variable.
> V. Instance variable not being used globally within the class. i.e., an instance variable can be reduced to a local variable.
> W. Switch statement does not contain a break.
> X. Switch statement without default case.
> Y. Out of array bounds by using <= instead of <.

*Figure 2: Hristova's rules for novice programmers*

## 3. Rules for Novice Programmers

A list of 20 errors that are often made by beginning programmers was created and discussed by (Hristova, 2003). These have been gathered by talking to experts and noting what they have experienced while teaching students how to program.

We classified these rules into four categories, with respect to PMD:
1. Errors that cause the program to be unparsable, and thus not fit for static analysis.
2. Errors currently not found by PMD.
3. Errors found by PMD but with feedback not suitable for novice programmers.
4. Errors found by PMD with suitable feedback for novice programmers.

Given the list of rules in Figure 2 we found that rules C, F, G, H, and L yield unparsable code, errors A, D, J, N, S, T, U, V and Y are not found, B, E, W, and X are found but with unsuitable feedback, and errors I, K, M, O, P, Q, and R are found with suitable feedback. PMD has an option to implement custom made rules and rulesets. A ruleset "Novice" was created with custom rules to cover errors that are missing or have unsuitable feedback.

The warning created by the new rules consists of the following elements: First, the warning specifies what part of the code generates the error. This typically consists of a line number or, in Eclipse, a small arrow in the side column to indicate there is something wrong on this line and an explanation of the error.
Second is a suggestion on how to fix the error. This could be one suggestion, or multiple. For example, if an instance variable is only used in one method, one suggestion could be to make that variable local to that one method. However, another suggestion could be to make the variable 'final', to indicate it being a constant.

Last is a reference to information on the concept that is presumably poorly understood. If possible we tried to refer the errors back to the textbook "Programming and object oriented design using Java" by Nino and Hosch [5], as book is used in our main first year programming course. This was included if applicable.

We encountered a few challenges when extending PMD, mainly around creating rules that look at multiple classes or projects as a whole. In PMD, there is no method which is called after the entire class is analyzed using a certain rule. This makes it more complicated to add rules that build on information collected from the entire class. Similarly, it is difficult to use information from other classes, if that class has not yet been analyzed by PMD. The order in which classes are analyzed matter. For these reasons, we implemented only new rules for seven of the nine errors that were not sufficiently covered by PMD.

## 4.  Experimental Results

To evaluate the effectiveness of the new rule set in finding mistakes made by novice programmers, we analyzed code by novice and professional software projects. The novice projects are assignments and final projects from a first year programming course at our university. When students created these projects, they were near completion of the course. The professional code was taken from six parts of the '*org.apache.commons.codec*' library: *.codec* itself, *.binary*, *.digest*, *.language*, *language.bm* and *.net*. The results of this comparison are depicted in Table 1.

The results show that the new set of rules for novice programmers are effective in catching typical novice mistakes. Novice code has only a slightly increased number of warnings when we consider the standard set of rules used by PMD, an increase of about 17%. When we look in comparison at the warning by the novice set of rules we see an increase of 164% more warning in novice code.

*Table 1: Comparison of two code bases given the two sets of rules*

|  | Projects | Total Size in LoC | Standard warnings | Standard Warnings per 1kLoC | Novice warnings | Novice Warnings per 1 kLoC | Projects w/o novice warning |
|---|---|---|---|---|---|---|---|
| Novice code | 24 | 89056 | 59462 | 667.7 | 592 | 6.6 | 0 |
| Expert code | 6 | 6485 | 3679 | 567.3 | 16 | 2.5 | 4 |

*Table 2: Top 3 of warning in novice code*

| Rule | Description | Number of Warnings | Percentage of warnings |
|---|---|---|---|
| V | Instance variable not being used globally | 464 | 78.4% |
| W | Switch statement does not contain a break. | 34 | 5.7% |
| U | Many if/else checks on the same variable. | 33 | 5.6% |

The new rules are effective in finding mistakes that are indicative of novice programmers. This finding is also supported by the fact only two of the six professional project have any novice warning, while all 24 student projects have at least some potential mistakes. This confirms the observation that this are problems that experienced programmers rarely make.

If we look at the Top 3 of errors made by novice programmer in Table 2, we see that the most common errors relate to the structure of code. Apparently, novice programmers struggle with deciding the appropriate scope of variable, and the correct use of control structures.

## 5. CONCLUSIONS AND FUTURE WORK

This paper describes how to adapt the off-the-shelf static code analysis tool PMD to support teaching programming to novice students. It discussed how the fact that these tools were developed for a different purpose - namely to support software development - make them less than ideal. We identified the need to provide a different type of feedback to students; one that identifies misunderstood concepts, instead of errors that need to be fixed. We used a list of 20 common novice mistakes by Hristova (Hristova, 2003) to guide the development of a new set of rules for PMD. We compared its effectiveness in finding errors in novice code, by comparing the results for a novice student projects, with mature software projects.

The student project were by students at the end of an intensive programming course. This means that the code if fairly mature for student code. Future work will be to see how the tool can effectively incorporated throughout the course, and how it can help teaching assistants who play an important role in helping student to avoid such mistakes. We will also investigate how to broaden the scope of the analysis to better cover concepts that novice students often misunderstand.